\documentclass[11pt,draft]{article}
\usepackage{amsmath, amssymb, amsthm, amsfonts, latexsym, color, paralist, cancel, ifdraft}
\usepackage[final]{graphicx,hyperref}
\usepackage{authblk}

\newlength{\actualtopmargin}
\newlength{\actualsidemargin}
  \theoremstyle{plain}
  \newtheorem{theorem}{Theorem}
  \newtheorem{lemma}[theorem]{Lemma}
  \newtheorem{corollary}[theorem]{Corollary}

  \theoremstyle{definition}
  \newtheorem{definition}[theorem]{Definition}

  \theoremstyle{remark}
  
  \theoremstyle{plain}
  \newtheorem*{theorem*}{Theorem}
  \newtheorem*{lemma*}{Lemma}
  \newtheorem*{corollary*}{Corollary}
  \newtheorem*{proposition*}{Proposition}
  \newtheorem*{claim*}{Claim}
  
  \DeclareMathOperator*{\argmin}{arg\,min}

\newcommand{\ii}{\mathbb{I}}

\newcommand{\norm}[1]{\left\| #1 \right\|}
\newcommand{\bra}[1]{\langle #1 \vert}

\newcommand{\ket}[1]{\vert #1 \rangle}
\newcommand{\tr}{Tr}

\newcommand{\ketbra}[1]{\vert #1 \rangle \langle #1 \vert}
\newcommand{\braket}[2]{\langle #1 \vert #2 \rangle}

\newcommand{\tensor}{\otimes}

\newcommand{\vvv}[2]{\left[ \begin{array}{l} #1 \\ #2 \end{array}\right]}

\newcommand{\mmm}[4]{\left[ \begin{array}{cc} #1 & #2 \\ #3 & #4\end{array}\right]}

\begin{document}
\title{\Large \textbf{An adaptive attack on Wiesner's quantum money}}
\author[1]{Daniel Nagaj}
\author[2]{Or Sattath}
\author[3]{Aharon Brodutch}
\author[4]{Dominique Unruh}
\affil[1]{Institute of Physics, Slovak Academy of Sciences}
\affil[2]{Computer Science Division, UC Berkeley}
\affil[3]{Institute for Quantum Computing and Department of Physics and Astronomy, University of Waterloo}
\affil[4]{University of Tartu, Estonia}
\renewcommand\Authands{ and }
\maketitle
\vspace{-5mm}

\begin{abstract}

Unlike classical money, which is hard to forge for practical reasons (e.g. producing paper with a certain property), quantum money is attractive because its security might be based on the no-cloning theorem. 
The first quantum money scheme was introduced by Wiesner circa 1970. Although more sophisticated quantum money schemes were proposed, Wiesner's scheme remained appealing because it is both conceptually clean as well as relatively easy to implement.

We show efficient adaptive attacks on Wiesner's quantum money scheme~\cite{wiesner1983conjugate} (and its variant by Bennett et al.~\cite{bennett1983quantum}),
when valid money is accepted and passed on, while invalid money is destroyed.
We propose two attacks, the first is inspired by the Elitzur-Vaidman bomb testing problem~\cite{elitzur1993quantum,kwiat1995interaction}, while the second is based on the idea of {\it protective measurements}~\cite{aharonov1993meaning}. 
It allows us to break Wiesner's scheme with 4 possible states per qubit, and generalizations which use more than 4 states per qubit.  
The attack shows that Wiesner's scheme can only be safe if the bank replaces valid notes after validation.
\end{abstract}
\section{Introduction}


One of the main requirements for any medium of money is that it should not be easily copied. For this very reason, it is appealing to construct {\em quantum money}: its security would follow from the laws of quantum mechanics, or more specifically, the no-cloning theorem~\cite{wootters1982single}. Indeed, quantum money was one of the earliest quantum information protocols, introduced by Stephen Wiesner circa 1970, although it took some time to be published~\cite{wiesner1983conjugate}.

Wiesner's quantum money scheme uses only single-qubit memory and single-qubit measurements, as follows: A bank creates a note of size $n$ with a public serial number $s$, and for each serial number a random (classical) private key $k^{(s)}\in \{0,1,+,-\}^{n}$. The corresponding banknote contains a {\em quantum money state} $\ket{\$_{s}}=\ket{k^{(s)}_{1}}\tensor \ket{k^{(s)}_{2}}\tensor \ldots \tensor \ket{k^{(s)}_{n}}$, where $\ket{+}=\frac{1}{\sqrt{2}}(\ket{0}+\ket{1})$ and $\ket{-}=\frac{1}{\sqrt{2}}(\ket{0}-\ket{1})$.
The serial number together with the quantum money state, i.e. the pair $\left(s,\ket{\$_{s}}\right)$, form {\em legitimate} quantum money. 

In order to validate her money (and pay with it), Alice sends the (potentially forged) banknote $(s,\ket{\psi})$ to the bank. The bank measures each of the qubits of $\ket{\psi}$ in its respective basis; the $i^{th}$ qubit is measured in the basis $\{\ket{0},\ket{1}\}$ if $k^{(s)}_{i} \in \{0,1\}$, and in the $\{ \ket{+}, \ket{-} \}$ basis otherwise. The money is declared valid if and only if all the measurement outcomes agree with the measurements on the legal state $\ket{\$_s}$. 

One downside of Wiesner's original scheme is that the bank must keep a database containing the secret key for every serial number. In a follow-up paper, Bennett, Brassard, Breidbart and Wiesner used a fixed pseudo-random function for choosing the secret keys for all the serial numbers, which implies that the memory required by the bank does not grow as a function of the number of legitimate serial numbers~\cite{bennett1983quantum}. All our results apply both to Wiesner's and to the Bennett et al. scheme.   

There are two specifications which are important for our work: we have to determine whether after a successful validity test, the money state is returned to Alice (or passed to Bob who does business with Alice), or replaced with a new quantum money state and a new serial number. Next, after a failed validity test, is the post-measurement state (a bad bill) returned to Alice for inspection? These distinctions are crucial for the scheme's security. We define \emph{strict testing} to be the variant of Wiesner's scheme in which the valid money state is returned to the owner after a successful test, and the post measurement state of a failed test is destroyed. 

Wiesner proposed his quantum money scheme more than 40 years ago,  and it was believed to be secure although a complete security proof was never published. However, Lutomirski~\cite{lutomirski2010online} and Aaronson~\cite{aaronson2009quantum} independently observed that the scheme is insecure if the bank returns valid notes as well as the post measurement state after detecting an invalid note. Farhi et al. showed that ``single copy tomography''~\cite{farhi2010quantum} can be performed in a much more general setting. When we own a single copy of an unknown state $\ket{\psi}$ and have access to a projective measurement (validation test) $\left\{\ket{\psi}\bra{\psi}, \ii-\ket{\psi}\bra{\psi}\right\}$ provided as a black box, we can efficiently estimate the reduced local density matrices of the state $\ket{\psi}$. This approach, based on Jordan's lemma, pointed out the security threat posed by having access to a validation procedure (in particular, the post-measurement state). Lutomirski's conclusion was that as long as the bank does not return the post-measurement state of an invalid note, the scheme should remain safe. 

On the other hand, Molina, Watrous and Vidick proved that Wiesner's scheme is secure against \emph{simple counterfeiting attacks}~\cite{molina2013optimal}. In this model, an attacker is given a single copy of an authenticated state, and attempts to create two banknotes with the same serial number which, independently, pass the bank's validity test. During the counterfeiting process, the attacker does not have access to the validity test. They showed that the success probability of the optimal attack on Wiesner's quantum money scheme is $\left(\frac{3}{4}\right)^{n}$.

Patawski et al.~\cite[Theorem 5]{Pastawski02102012} proved security under a more general setting. In their setting a user is given one bank note, and creates polynomially many faked banknotes. The bank validates all the banknotes. They show that the probability that at least two banknotes would pass the validation is exponentially small. However, in a general attack (as in our case), the attacker could make some use of the quantum validation procedure, even if the bank strictly discourages failed tests and does not return bad banknotes.

\paragraph{Main results.} 
We show that in a \emph{strict testing} variant of Wiesner's scheme (that is, if only valid money is returned to the owner), given a single valid quantum money state $(s, \ket{\$_s})$, a counterfeiter can efficiently create as many copies of $\ket{\$_s}$ as he wishes (hence, the scheme is insecure). 
He can rely on the quantum Zeno effect for protection -- if he disturbs the quantum money state only slightly, the bill is likely to be projected back to the original state after a test. Interestingly, this allows a counterfeiter to distinguish the four different qubit states with an arbitrarily small probability of being caught.

{\bf The BT (bomb-testing) attack:} The simplest attack is based on the Zeno assisted Elitzur-Vaidman bomb test \cite{kwiat1995interaction}. The \emph{bomb testing} (BT) attack lets us tell whether the first qubit of our quantum money is in the state $\ket{+}$ or not. By repeating this test for each of the money qubits (and for each of the four possible states), we can identify the quantum money state. We use an ancillary {\it probe} qubit, initialized to the $\ket{0}$ state. We repeat the following steps $N=\frac{\pi}{2\delta}$ times, as depicted in Fig.~\ref{fig:attack} on p.~\pageref{fig:attack}: 
\begin{enumerate}
\item Rotate the probe by a small angle $\delta$. 
\item Apply a C-NOT from the probe qubit to the money qubit. 
\item Send the quantum money to the bank for validation (and get it back when verified).
\end{enumerate}
If the money qubit is in the $\ket{+}$ state, it stays invariant under the NOT operation, and therefore also by the C-NOT operation controlled by a probe. Hence, at the end of the procedure, the probe qubit will be in the state $\ket{1}$. If the quantum money state is in either the $\ket{0}$ or $\ket{1}$ state, the probe will be in the $\ket{0}$ state, using the same analysis of the Elitzur-Vaidman bomb tester. In these two cases the maximal rotation induced on the money state at any one time is at most $\delta$ and the bank's probability of detecting a counterfeiter is at most $\delta^2$; hence, the overall probability of detection by the bank is $O(\delta)$. The last case, $\ket{\$_i}=\ket{-}$,  is somewhat different: after the first iteration, the probe has angle $-\delta$. At the end of the second iteration, the probe returns to state $\ket{0}$, etc. Therefore, at the end of the procedure, the probe is in the state $\ket{0}$ with certainty
(as long as $N$ is even), while the money state is left invariant.

One might hope that a simple generalization of Wiesner's strict testing scheme using $r$ states as a basis instead of $4$ states (and a $d$ dimensional qudit instead of a qubit) will be able to hold off our attack. However, we show that this is not sufficient. More precisely, let the generalized money scheme use $n$  random states from the set $\{ \ket{\beta_1},\ldots,\ket{\beta_r}\}$, where $\ket{\beta_i} \in \mathcal{C}^d$, and let 
\[\theta_{min}= \min_{ 1\leq i \neq j \leq r}  \arccos|\braket{\beta_i}{\beta_j} |.\]
In Appendix~\ref{testLIST} we show a simple generalization of the BT attack that succeeds with probability $1-f$, and uses 
$O\left(n r^2 \theta_{min}^{-2}f^{-1}\right)$ validations. 

{\bf The PM (protective measurement) attack:} What if there are infinitely many states per qubit
(which implies an infinite number of verifications, since $\theta_{min} = 0$)? 
In this case, the previous attack fails (see Appendix~\ref{bomb_testing_fails}), and we present an alternative tomographic attack based on protective measurements~\cite{aharonov1993meaning}. The \emph{protective measurement} (PM) attack allows us to estimate the expectation value $\langle A\rangle = \bra{\psi}A\ket{\psi}$ of an operator $A$ in the state $\ket\psi$, without disturbing the state much, by preparing a probe in the initial state $\ket{0}$, choosing $\delta = \frac{c}{N}$ for some constant $c$ and repeating the following procedure $N$ times
\begin{enumerate}
\item Weakly couple the probe and the system.  
\item Send the state to the bank for validation.
\end{enumerate}
We aim for the following to happen:
\begin{align*}
	\ket{0}\ket{\psi}\, 
		&\xrightarrow{\mbox{\Large{$e^{-i\delta \left(\sigma_x\tensor A\right)}$ }}} \, \approx \,
	\ket{0}\ket{\psi} - i \delta \ket{1} A \ket{\psi}  \\
		& \xrightarrow{\text{bank measures } \{\ketbra{\psi},\, \ii-\ketbra{\psi}\}} \, \approx \,
	\left(e^{-i\delta \langle A\rangle \sigma_x} \ket{0} \right) \otimes \ket{\psi} \\
	& \xrightarrow{\text{repeat $N$ times}} \,\approx \, 
	\left(e^{-iN\delta \langle A\rangle \sigma_x} \ket{0} \right)
		\otimes \ket{\psi}.
\end{align*}
By measuring the probe and using standard parameter estimation techniques, we can approximate $\langle A \rangle$ and thus $\ket{\psi}$ to the desired precision (depending on $n$, the number of qubits in the unknown state $\ket{\psi}$). 

In Def.~\ref{def:protective_tomography} we present  a more general question regarding the cost,  accuracy and confidence of \emph{protective tomography}, i.e. where we have a single copy of a state $\ket{\psi}$ and access to a validation procedure $\{\ket{\psi}\bra{\psi},1-\ket{\psi}\bra{\psi}\}$ that destroys the state when the validation fails. As far as we know, our analysis in 
Sec.~\ref{sec:protective_measurement}, gives the first quantitative answer to this question.

Using this tomography approach, without any assumptions on the states, for any constant $\epsilon$, using $O\left(n^5 \ln^2(n)\right)$ bank validations, we get caught with probability $1-\epsilon$; and conditioned on the event that we are not caught, we find a (classical description of a) quantum state $\rho$ such that $F(\rho, \ketbra{\$}) \geq 1 -\epsilon$, with probability at least $1-\epsilon$. Note that $1-F^{2}(\rho,\ketbra{\$})$ is precisely the probability of getting caught when providing a fake state $\rho$ instead of the legitimate money $\ket{\$}$.

\paragraph{Discussion and applications.} Our attack applies in the strict-testing regime, where good banknotes are returned or passed on, while failed tests result in confiscation of the banknote (or us being sent to jail).
However, the attack does not work if after a valid test, a new quantum bill with a new serial number are returned to the owner. Does this affect the advantages of Wiesner's scheme? In order to answer this question, we first need to understand these advantages.

The two main advantages of Wiesner's scheme over other money schemes are the following (see a more detailed analysis in Ref.~\cite{gavinsky2012quantum}):
 \begin{itemize}
 \item The data needed for validation is static and classical. 
 Therefore, after the money has been issued, many bank branches can validate the money without any need for communication.
 \item The hardware requirements (single-qubit memory and single-qubit measurements) are less demanding than other schemes such as Farhi et al.~\cite{farhi2012quantum} and Aaronson and Christiano~\cite{aaronson2012quantum} which require a fully scalable quantum computer and quantum memory which can hold a large number of entangled qubits.
 \end{itemize}
These advantages remain when the quantum money is replaced with a new one after each successful validity test. In some sense, such quantum money resembles one time tokens, which are destroyed after each usage. 

Where do our results lead? First, we believe that the greatest potential of this work is in the context of weak measurements. The framework of weak measurement has been proven an important concept in numerous cases~\cite{Dressel2014, Aharonov2007} including protective measurements and precision metrology. Our ``bomb" attack shares a lot of the properties of weak measurements, but not all. We believe that further investigation  of these different approaches in a broad perspective will yield  practical applications for weak measurement methods.
Second, our adaptive attack is interesting on its own as a pedagogical device for exhibiting the counter-intuitive properties of quantum mechanics, and more specifically, for weak \& protective measurements. Third, it raises an important warning about quantum money constructions -- we need to be cautious reusing valid bills (even though false bills are destroyed).
Fourth, one can think about a failed test simply as an undesirable scenario, and try to apply the technique even if no actual strict-validation box exists. This could find use in tomography and state discrimination using several copies of unknown states. 
Fifth, after finishing this work we learned that our preprint and the strict-testing model has inspired interesting questions in query complexity \cite{BombQueryComplexity}.

\paragraph{Wiesner's money in a noisy environment.}


In this paper, we have focused on Wiesner's money in a noiseless
environment. That is, the bank rejects the money if even a single
qubit is measured incorrectly. In a more realistic setting, we have to
deal with noise, and the bank would want to tolerate a limited amount
of errors in the quantum state \cite{Pastawski02102012}, say 10\,\%. (All concrete numbers in
this section are examples.) Additionally, the natural design choice
would be that the bank repairs the incorrect qubits so that the
quantum state does not deteriorate more and more. But then, a
simple attack exists~\cite{lutomirski2010online,aaronson2009quantum}: The owner replaces the first qubit
of the money state by $\ket0$, keeping the original qubit. Then he
submits the money state for validation. The bank repairs the
replaced qubit. Now the user has a copy of the first qubit, and the
original money state. By repeating this process, the user may get
copies of all qubits, thus getting two copies of the money state. In
light of this attack, it seems obvious that the bank must issue a
fresh, independent money state (with a new serial number) even in case
of a successful validation. This is, of course, the same
recommendation as we are making in this paper: ``Issue a new money state
after each validation.'' 

We are aware of the following possible criticism of our
result, saying we are attacking an obvious loophole: ``Since in the case of noise, there are
obvious reasons why the bank should re-issue the money,
this attack is not relevant since the bank would prepare new money states anyway.'' 
However, this is not the case, our attack is relevant even in a setting with noise. 
The bank surely should not simply repair the incorrectly measured, noisy qubits. 
However, it is not strictly necessary to
always re-issue a new state either. Instead, the bank could do the following:
\begin{compactitem}
\item If at most $5\%$ of the qubits were measured incorrectly, the
  bank hands back the state. (No repairs.)
\item If $5\%$--$10\%$ of the qubits were measured incorrectly, the bank
  re-issues a new money state.
\item If more than $10\%$ of the qubits were measured incorrectly, the
  bank informs the police of a forgery attempt.
\end{compactitem}
Why would this be a reasonable thing to do? First, this approach takes
care of the degradation of the quantum money by re-issuing the money
when the error exceeds $5\%$. However, it also saves resources: the
money is not re-issued upon each validation. Hence, fewer serial
numbers need to be allocated, reducing the storage needed by the
bank.\footnote{This becomes particularly relevant when we do not use a pseudorandom function to map serial numbers to money states~\cite{bennett1983quantum} (because we want information-theoretical security) and we distribute the whole database with the serial numbers up front to the branches of the bank (because we want to allow the different branches to validate money independently without communication).}  Finally, it is no longer possible for the attacker to just replace
qubits to get a second copy of the money state. In fact, none of the
prior attacks we know of applies to this scheme. The attack proposed
in this paper, however, applies to this scheme without any
modification. Thus our attack does indeed constitute a new attack
vector, and needs to be taken into account in the design of quantum
money protocols (and possibly other quantum protocols). We stress that
the above protocol example is just that -- an illustrative
example. Its purpose is to illustrate that there can be many settings
in which our attack applies, the above is just the simplest one we
could come up with.


\paragraph{Structure of the paper.} We introduce the quantum (Zeno effect based) 
Elitzur-Vaidman bomb quality tester in Section~\ref{sec:EVbomb} and use it as a tool to enlighten the rest of the analysis in Section~\ref{sec:attack}, where we show our adaptive BT (bomb-testing) attack on Wiesner's scheme. 
We present and analyze the PM (protective measurement) attack in Section~\ref{sec:protective_measurement} which lets us perform single-copy tomography with the help of strict testing, for general states, or for 
a generalization of Wiesner's scheme which was mentioned before.
Finally in Section~\ref{sec:discussion}, we briefly compare the two attacks. In  Appendix~\ref{sec:general} we extend the BT attack to deal with simple generalizations of Wiesner's scheme. However, in Appendix~\ref{bomb_testing_fails} we show that unlike the PM attack, the BT attack does not always work 
-- and it is instructive to learn why.  

%
%


\section{Elitzur-Vaidman's bomb quality tester}
\label{sec:EVbomb}
We usually think that in order to measure a quantum system we must interact with it. 
However, sometimes there is a possibility of an {\em interaction-free} measurement detecting 
some property of a system without disturbing it.
The prime example of this is Elitzur-Vaidman's bomb tester \cite{elitzur1993quantum}, 
a probabilistic test that can certify a property of an object (a working trigger) 
by detecting a photon that never ``interacted'' with the object.
Using the quantum Zeno effect, this test has been improved \cite{kwiat1995interaction} 
so that one can be sure about the system's property, while the probability 
of disturbing the object goes to zero (we do not want any explosions).


We now present a quantum information variant of this approach, which we will use in the next section as an adaptive attack against Wiesner's quantum money scheme.

\begin{figure}
\begin{center}
\includegraphics[width=12cm]{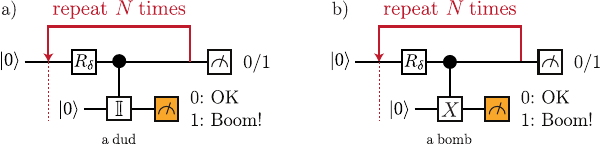}%
\caption{A quality-testing procedure for bombs: run $N$ rounds and end with a measurement of the first register. a) A dud can't explode, and the first register slowly rotates from $\ket{0}$ to $\ket{1}$. b) With a live bomb, we can really trigger the bomb by flipping the second register to $\ket{1}$. This does not happen often as $\delta$ is small, and we are much more likely to measure $\ket{0}$ on the second register. The first register is then projected back to $\ket{0}$.}%
\label{fig:bomb}%
\end{center}
\end{figure}

The goal is to test whether a ``quantum bomb'' is a dud or an actual bomb\footnote{
Note that we differ somewhat from the original treatment of the bomb-tester, as our trigger is a controlled quantum gate ($I$ or $X$ in Figure~\ref{fig:bomb}) instead of a transparent object for a dud and a photon detector for a working bomb. However, mathematically, our presentation is equivalent to the original one, and it turns out to be more convenient for understanding Section~\ref{sec:attack}.
}. 
If we have a dud, it remains in the $\ket{0}$ state when we interact with it.
On the other hand, we can flip the state of a live bomb to $\ket{1}$, which makes it explode.
The trick is how not to trigger a live bomb. The safe bomb quality testing procedure is illustrated in Figure~\ref{fig:bomb}. 
We pick a large number $N$, choose a small angle $\delta$ and label $R_\delta$ a counterclockwise rotation by this angle:
\begin{align}
	\delta = \frac{\pi}{2N}, \qquad
	R_\delta = \mmm{\cos \delta}{-\sin \delta}{\sin \delta}{\cos \delta}. \label{delta}
\end{align}
We will use two registers (the probe and the system), and apply the {\em controlled interaction}
\begin{align}
	C_P = \ket{0}\bra{0} \otimes \ii + \ket{1}\bra{1} \otimes P. \label{probe}
\end{align}
When there is an active bomb, this operation is a controlled-$X$ (a CNOT), while for a dud $P=\ii$, so the operation $C_P$ is just an identity.
The testing procedure starts with the first register in the state $\ket{0}$ and applies the following steps $N$ times
\begin{enumerate}
	\item Prepare the second (system) register in $\ket{0}$.
	\item Rotate the first (probe) register using $R_{\delta}$.
	\item Apply the controlled interaction $C_P$. 
	\item Measure the second (system) register. If we get $\ket{1}$, we fail. If we succeed, we reuse the first register and go back to step 1.
\end{enumerate}
Once we are done with $N$ repetitions of these steps, we measure the first register in the computational basis.

First, let us look at the case with a dud (Figure~\ref{fig:bomb}a).
The controlled interaction does nothing, and the state after the first 
round is $\left(\cos \delta \ket{0}+\sin \delta \ket{1}\right)\ket{0}$. We thus measure $\ket{0}$ in the second register, and continue. In the following rounds, the first register is repeatedly rotated by $\delta$, finally ending up in the state $\ket{1}$. Thus, without a bomb, we finally end up measuring $\ket{1}$ in the first register.

Second, what if there is an actual bomb? Now $P=X$, and the controlled interaction is the CNOT operation. If we actually tested the bomb (flipped the second register), we would die (measure $\ket{1}$ in the second register). 
However, we choose to test the bomb in superposition, with a small angle $\delta$ of the control register state.
In the first round, the state before the measurement is $\cos \delta \ket{0}\ket{0}+\sin \delta \ket{1} \ket{1}$, and the probability of an explosion is $\sin^2 \delta$. 
Detecting an explosion is a measurement. Thus, if nothing is heard, we project both registers back to $\ket{0}\ket{0}$. 
The state of the system after a successful round is just what it was in the beginning! 
If the bomb never explodes, the control register remains in the state $\ket{0}$ throughout the $N$ rounds. 
The probability of getting no explosion in these $N$ steps is
\begin{align}
	 \left(1 - \sin^2 \delta \right)^N 
	\geq 
	\left(1-\frac{\pi^2}{4N^2}\right)^N 
	\geq
	1 - N\frac{\pi^2}{4N^2}
	=1 - \frac{\pi^2}{4N} .  \label{pliveEV}
\end{align}
Thus, we will live through the $N$ steps and measure $\ket{0}$ in the first register with probability approaching 1, and conclude that there is a ``live'' bomb in the system. 

To conclude, we can ``safely'' discern the quality of a ``quantum bomb'' by relying on the quantum Zeno effect (in other words, because a watched quantum pot never boils). Things get more interesting when we apply this test to other input states besides $\ket{0}$ in the second register. In the next section we show this results in a successful attack against Wiesner's quantum money scheme.
\section{An adaptive attack on Wiesner's quantum money}
\label{sec:attack}

\begin{figure}
\begin{center}
\includegraphics[width=3.5cm]{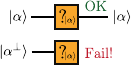}%
\caption{A strict (unforgiving) money tester accepts a valid state and hands it back to us (or to whom we pay). However, if we hand in an orthogonal state, we fail and rot in jail.}%
\label{fig:trap}%
\end{center}
\end{figure}

We now show that if validated banknotes are forwarded to a new owner (or returned to the original owner), Wiesner's quantum money is vulnerable to an adaptive attack.
The attack works also when the bank destroys bad bills, or sends us to jail if we try to validate a bill that does not pass the test (see Figure~\ref{fig:trap}).\footnote{As mentioned in the introduction, the scenario where the bank returns the bad bills, that is, when the attacker has access to the post-measurement result in failed cases, was already analyzed and successfully attacked by Lutomirski \cite{lutomirski2010online}, Aaronson~\cite{aaronson2009quantum}, and Farhi et al.~\cite{farhi2010quantum}.}

Failing a validation test is undesirable, analogous to a ``bomb explosion'' from the previous Section. Motivated by the success of Elitzur-Vaidman's bomb tester, we will try to extract information about our system (learn in which of the 4 states the $i^{\text{th}}$ qubit of the quantum money state is in) without triggering the ``bomb'' (being sent to jail).

\begin{figure}
\begin{center}
\includegraphics[width=13cm]{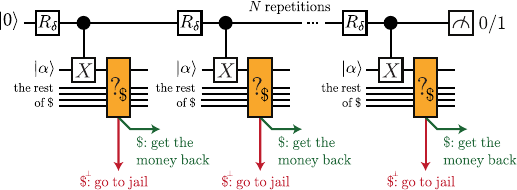}%
\caption{An adaptive attack on Wiesner's quantum money with a strict testing procedure. We can identify whether the qubit $\ket{\alpha}$ is in the state $\ket{+}$ without going to jail (being detected). If we do not identify it, we can use controlled-$(-X)$ instead to test for $\ket{-}$. If we do not detect it either, we just measure the qubit in the computational basis.}%
\label{fig:attack}%
\end{center}
\end{figure}

Let us have a single valid banknote with the $n$-qubit state $\ket{\alpha_1}\tensor\ket{\alpha_2}\tensor\cdots\tensor \ket{\alpha_n}$ with $\ket{\alpha_i}\in \{\ket{0},\ket{1},\ket{+},\ket{-}\}$.
We show how to determine any particular single qubit state $\ket{\alpha}$ from the banknote by following the attack in  Figure~\ref{fig:attack}. 
There are two differences from Figure~\ref{fig:bomb}.
First, the input state $\ket{\alpha}$ is unknown -- a qubit from the banknote can be any one of the four possible states $\ket{0}, \ket{1}, \ket{+}, \ket{-}$. Second, instead of doing measurements in the $\ket{0},\ket{1}$ basis on the second register, the bank measures the qubit in the (unknown to us) $\ket{\alpha},\ket{\alpha^\perp}$ basis, verifying whether we gave it the state $\ket{\alpha}$ and the rest of the undisturbed banknote qubits. The test is unforgiving, sending us to jail if the money tests false, i.e. when the unknown qubit projects to $\ket{\alpha^\perp}$.

What happens in Figure~\ref{fig:attack} when we flip the four possible qubit states using an $X$ operation?
\begin{enumerate}
	\item First, we analyze the case that the unknown state is $\ket{0}$ or $\ket{1}$.  Flipping maps the states $\ket{0} \leftrightarrow \ket{1}$ to each other, and  the validity tester rejects the flipped state. Thus, we can view these two qubit states as the ``live bomb'' case in Figure~\ref{fig:bomb}b). Measuring the validity of the bill will result in projecting the control qubit back to $\ket{0}$ every time. In the end, the control (first register) will remain $\ket{0}$, and we will pass through the procedure with probability arbitrarily close to one \eqref{pliveEV}.
	\item Second, when the unknown qubit is $\ket{+}$, a flip does nothing to it. Thus, it behaves like the case with a ``dud'' in Figure~\ref{fig:bomb}a). In Figure~\ref{fig:attack}, this results in a rotation of the the control qubit to $\ket{1}$ in $N$ steps. We are never caught doing anything illegal in this case.
	\item Finally, when the input state is $\ket{-}$, a bit flip gives it a minus sign, 
	which makes things more interesting. The initial state is $\ket{0}\ket{-}$. We apply the rotation $R_\delta$ and a CNOT, obtaining
\begin{align}
	R_\delta \otimes \ii: & \,\, 
			\left( (\cos \delta) \ket{0} + (\sin \delta) \ket{1}\right) \ket{-}, \\
	\textrm{CNOT}: & \,\,
			\left( (\cos \delta) \ket{0} - (\sin \delta) \ket{1}\right) \ket{-}.
\end{align}
The quantum money still passes the test perfectly. 
However, the relative sign of the first register state is now negative, and its angle with the $\ket{0}$ state is $-\delta$.
Let us do the second iteration. We rotate the first register from $-\delta$ to $\ket{0}$, and the following CNOT does not do anything. The third round is just like the first round, the fourth like the second, etc. After an even number of rounds, the state of the first register will be $\ket{0}$. Meanwhile, all the tests will have passed perfectly, and we are without any danger of being caught.
\end{enumerate}

Therefore, if the qubit $\ket{\alpha}$ was in the $\ket{+}$ state, we can identify it accurately, with impunity. How to identify the other three cases?
We can test for the state $\ket{-}$ using the controlled-$(-X)$ operation in Figure~\ref{fig:attack}. Hence, we can rule out (or certify) $\ket{-}$ as well, conclude that $\ket{\alpha} \in  \{\ket{0}, \ket{1}\}$, and safely measure it in the computational basis.

Wiesner's strict testing (good money returning, bad money confiscating), 4-state $\{\ket{0},\ket{1},\ket{+},\ket{-}\}$ scheme is thus vulnerable to an adaptive attack. 
Suppose we want our attack to fail with probability at most  $f$ for $f \ll 1$. For a bill with $n$ secret qubits, we choose $N=\frac{\pi^2 n}{2 f}$, and run the procedure in Figure 3 at most twice per qubit -- once for every qubit $i$ in order to distinguish whether the $i^{\text{th}}$ qubit is in the state $\ket{+}$, and if it was not, run the alternative test to distinguish whether it is in the state $\ket{-}$. If it isn't -- the state is a computational basis state, and we can now safely measure it. Thus, $2N$ verifications are necessary to identify a single qubit. For each run of the $N$-step procedure in Figure 3, the probability of failing is at most
$\frac{\pi^2}{4N}$, according to Eq.~\eqref{pliveEV}. Therefore,
\begin{align}
	\Pr(\text{attack succeeds}) \geq \left(1-\frac{\pi^2}{4N}\right)^{2n} 
	\geq 1 - \frac{\pi^2 n }{2N}=1-f,
	\label{attack_succeeds}
\end{align}
Thus, we can identify all $n$ qubits using at most $2n \times N$
adaptive questions to a strict (unforgiving) tester. 

However, we can submit a banknote for verification with all $n$ qubits slightly changed (in parallel, with one probe qubit per system qubit). This way, each verification fails with probability $O\left(n \delta^2\right)$,
and running $N$ successful rounds fails with probability $O\left(N n \delta^2\right) = O\left(n N^{-1}\right)$. 
Thus, if we set $f$ for our final failure probability, it is enough to use $N = O\left(nf^{-1}\right)$,
which translates to only
 \begin{align}
	2N =  \frac{\pi^2 n}{f}
    \label{eq:verifications_in_parallel}
\end{align}
verification rounds, saving a factor of $n$.



We can try to salvage Wiesner's strict testing money scheme from this attack by adding other states for the banknote qubits. Adding the $y$-basis states does not help, as the CNOT flips them to orthogonal states, and they correspond to a ``live bomb'' case again. To detect them, we would perform tests with a controlled-$Y$ or controlled-$(-Y)$ instead of a CNOT.
We could also try to use a collection of unrelated single-qubit states.  If the list of possible states is finite (and known), we can use a  similar procedure, outlined in Appendix~\ref{sec:general}. However, we can also deal with the most general case (where the list of possible states is infinite, or is unknown) as shown in the next Section. 




\section{Another way to attack: 
a protective measurement}
\label{sec:protective_measurement}

In the BT attack above, the quantum Zeno effect kept us safe from explosions (our attack being detected by the bank). It worked nicely because of the special relationship between the four states $\ket{0},\ket{1},\ket{+},\ket{-}$. They were flipped
or kept intact. However, what would happen if we analyzed a different list of single-qubit quantum money states, on which the CNOT operation did something else? The behavior of the probe system and the probabilities of failure are calculated in detail in Appendix~\ref{sec:general}. 
We see there that if two of the possible states are very close to each other, the BT attack is not weak enough, and we cannot get a satisfactory upper bound on the cumulative probability of failure.
We now present a different attack which ensures that we are safe enough in any round (and in summary, in an $N$-round procedure). Below we show that it is possible to construct an approximate state $\rho$ such that the fidelity\footnote{The definition of the fidelity is 
$F(\rho,\sigma) = \textrm{Tr}\left(\sqrt{\rho^{1/2}\sigma \rho^{1/2}}\right)$.
It can be easily seen that if one of the two states is pure, the fidelity satisfies:
$F(\rho,\ket{\psi})= \sqrt{\bra{\psi} \rho \ket{\psi}}$.
} $F(\rho,\ket{\alpha})>1-\epsilon$ and estimate the running time, confidence levels and probability of success  for the procedure. The fidelity squared gives the probability that the bank will accept our counterfeit $\rho$ as valid.

The basic building block of this method is to ensure {\em weak interaction} between the probe and the system.
The method is reminiscent of the {\em quantum random walk} and {\em protective measurement} ideas of~\cite{aharonov1993quantum,aharonov1993measurement}.
We let a probe system interact weakly with the bill at each step while maintaining coherence of the bill state. The probe state will evolve as a linear function of the weakness parameter $\delta$, while the probability of a failed validation will be quadratic in $\delta$. The procedure is called a protective measurement since the validation step protects the money state by projecting it back to its original state with high probability. 

 The protective measurement scheme was originally derived as a method to fully measure  the wavefunction of an unknown protected quantum state (essentially performing tomography) without disturbing the state. The probe is usually a continuous variable that can be used to record expectation values with the desired accuracy. Since the motivation was conceptual rather than practical, there has never been any  attempt to quantify the resources required for full tomography using this scheme. Apart from our use of protective measurement in a practical scenario, we describe a protective measurement scheme with a qubit probe and bound its running  time.

First, we will describe how the validation
procedure of the bank can be used to estimate the expectation value of
any dichotomic observable (i.e., an observable with eigenvalues
$\pm 1$)
\begin{align}
	A = P-P^\perp, \label{Adef}
\end{align}
where $P$ is a projector on its $+1$ eigenspace (in this work, $A$ will always be one of the Pauli matrices),
and $P^\perp=I-P$.

The basic idea for estimating $\langle A \rangle$ is to use a weak
interaction between a probe in some state $\ket{\varphi_0}$ and the money
state. Measuring the money state will then affect the probe state to a
small degree.  By repeating the weak interaction and validation, we
finally (approximately) transform the probe state into:
\begin{align}\label{eq:approx.unitary}
	\ket{\varphi_N} \approx e^{-i c \langle A \rangle\sigma_x}\ket{\varphi_0},
\end{align}
for some constant $c$. The following is a formal description of the above intuition. 

\begin{definition}[Protective Measurement]
You are given a single copy of an unknown state $\ket{\alpha}\in \mathbb{C}^d$, and access to a two outcome von Neumann measurement $\{\Pi=\ketbra{\alpha} , I-\Pi\}$, the validation. We say that a protocol is a  protective measurement of a dichotomic observable $A$  with running time $N$, accuracy $\epsilon$, and failure probability $f$  when  (a) The protocol makes at most $N$ uses of the validation. (b) With probability that all the outcomes are $\Pi$ is at least $1-f$. In this case, the procedure maps $\ket{\varphi}\ket{\alpha} \rightarrow \left[ e^{-i \frac{\pi}{8} \langle A \rangle\sigma_x} \ket{\varphi} + O(\epsilon)\ket{\varphi'}\right]\ket{\alpha}$ for all $\ket{\varphi}\in \mathbb{C}^2$. 
\end{definition}
We note that this definition is slightly different then Aharonov and Vaidman's original definition \cite{aharonov1993measurement}. In particular they use continuous variables for the meter and consider other possible protection methods. 

We usually think of a measurement as a mapping which takes a quantum state to a probabilistic classical result (and perhaps the post-measurement state). A protective measurement, on the other hand, maps a quantum states to quantum state. 

\begin{lemma}\label{lemma:protmeas}
For any dichotomic observable $A$ there exists a protective measurement protocol with running time $N$, accuracy $O(1/N)$ and failure probability $O(1/N)$. 
\end{lemma}

We prove this Lemma in Sec.~\ref{sec:weak} below. The procedure used in the proof can be slightly modified to identify one of the four Weisner states (see Sec.~\ref{sec:simple}).

By generating many copies of $\ket{\varphi_N}$ and measuring in the $\sigma_y$ basis we get an estimate
of $\langle A\rangle$:

\begin{lemma}[Statistics from protective measurement]\label{le:sfp} For any $\nu , \eta, f > 0$, it is possible to use a protective measurement protocol  to estimate $\langle A \rangle$ with precision at least $\nu$, confidence at least $1-\eta$, probability of failure $O(f)$ and running time  $O\left(f^{-1}\nu^{-4}\ln^2(\eta^{-1})  \right)$.
\end{lemma}

The proof of this lemma in Sec.~\ref{sec:tomography} can be generalized to protective measurement protocols with a wider range of parameters.  

Estimating $\langle A\rangle$ will  allow us to perform tomography to get a classical description of the money state $\ket\alpha$ and ultimately produce its (approximate) copy -- and similarly for each unknown qubit. Formally, 

\begin{definition}[Protective Tomography]You are given a single copy of an unknown state $\ket{\alpha} \in \mathbb{C}^d$, and access to a two outcome von Neumann measurement $\{ \Pi=\ketbra{\alpha}, I - \Pi \}$  the validation. We say that a protocol achieves protective tomography with infidelity  $\epsilon$, confidence $1-\eta$, failure probability  $f$ and  running time $t$ if it outputs a classical description of a mixed state $\rho$ such that:
\begin{inparaenum}[(a)]  \item the probability of failure, i.e. that at some step of the algorithm the outcome of the measurement is $I-\Pi$, is $O(f)$.
\item if the algorithm does not fail, with probability at least $1-\eta$, we have $F(\ket{\alpha},\rho) \geq 1 - \epsilon$,
\item the algorithm uses at most $t$ validations.\end{inparaenum}

\label{def:protective_tomography}
\end{definition}

If the state $\ket{\alpha}$ is a product state of $n$ qubits, such as Wiesner's scheme and its extension (with infinitely many possible states per qubit, instead of 4), We can repeat the above procedure with adjusted parameters, and get an approximate classical description for $\ket{\alpha}$, i.e. 

\begin{theorem}\label{theorem:prottomo}
There exists a protective tomography protocol for a qubit system ($d=2$) with running time scaling as $t=O\left(f^{-1}\epsilon^{-4}\ln^2(\eta^{-1})  \right)$.
\end{theorem} 
We present the proof in Section~\ref{sec:P5}. Note that it can be easily extended to qudits of dimension $d$, with the running time now scaling as $t=O\left(d^{12} f^{-1}\epsilon^{-4}\ln^2(d^2 \eta^{-1})  \right)$. 
Next, we want to be able to perform protective tomography for a composite $n$-qubit product state. Theorem~\ref{theorem:prottomo} implies the following

\begin{corollary} There exists a protective tomography protocol for $n$-qubit states of the form $\ket{\alpha}=\bigotimes_{i=1}^n \ket{\alpha_i}$, with running time $t=O\left(n^5 f^{-1}\epsilon^{-4}\ln^2(n \eta^{-1})  \right)$.
\label{cor:protective_tomography_n_qubits}
\end{corollary}

We prove this Corollary in Section~\ref{sec:PTn}. In terms of the attack on Wiesner's quantum money, this is the final step producing the approximate classical description of $\ket{\alpha}$, with the probability of not being caught at least $1-f$. This in turn allows us to create as many counterfeit bills $\rho$ as we want. 
We also have a guarantee that with probability at least $1-\eta$, these counterfeit bills are going to be pretty good approximates of the original bill. Finally, if $\rho$ is a good approximation of $\ket{\alpha}$, then the first time each counterfeit bill is used, it can fail the bank's validation procedure with probability at most $2\epsilon$.

\subsection{Proof of lemma \ref{lemma:protmeas}: Coupling to the expectation value of a dichotomic observable}
\label{sec:weak}

\begin{figure}
\begin{center}
\includegraphics[width=8cm]{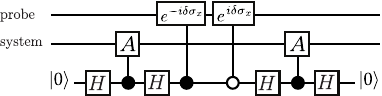}%
\caption{An implementation of $U$ from \eqref{Udefine}, involving projections on the $\pm 1$ eigenspaces of the observable $A$.}%
\label{fig:Uimplement}%
\end{center}
\end{figure}

The crucial difference from the approach of Section~\ref{sec:EVbomb} is that, instead of the rotation by $\delta$ and applying a CNOT to the probe/system, we use 
the unitary coupling operation
\begin{align}
	U &=
	e^{-i \delta \left(\sigma_x \otimes A\right)}
	= e^{-i \delta \left(\sigma_x \otimes P - \sigma_x \otimes P^\perp\right)} 
	= e^{-i \delta \sigma_x \otimes P} e^{i \delta \sigma_x \otimes P^\perp} 
	= e^{-i\delta\sigma_x} \otimes P + e^{i\delta\sigma_x} \otimes P^\perp, \label{Udefine} 
\end{align}
expressible in terms of the projector $P$. In this formula, we assume that system containing the probe is on the left, and the system containing the unknown state on the right of the tensor product.)
This works because we can divide the Hilbert space into two subspaces: the kernel and the range of $P$. These subspaces are invariant under the unitary $U$. Its action in each of the subspaces is then easily expressible.
Note that when $A$ is unitary with eigenvalues $\pm 1$, the unitary $U$ can be implemented as in Figure~\ref{fig:Uimplement}.

Choosing a small $\delta=\frac{c}{N}$ (and tuning the constant $c$ for optimal performance), we ensure that the interaction of the probe and the tested system is always weak -- independent of the relationship of $P$ and the unknown state $\ket{\alpha}$.
This is in contrast with the method in Section~\ref{sec:attack}, where if the probe was close to the state $\ket{1}$, the controlled interaction could change the state of the system a lot. 

We start with the probe qubit in an arbitrary  state $\ket{\varphi_0}$.
After $k$ rounds (assuming we have not yet been caught) the probe register will be in some state $\ket{\varphi_k}$ and the second register will hold the unknown state $\ket{\alpha}$. Let us apply $U$ again, and get the state $U\ket{\varphi_k}\ket{\alpha}$. This is followed by the validation, which gives us the unnormalized state 
\begin{align}
	W\ket{\varphi_k} = \left(\ii \otimes \bra{\alpha}  \right)U\ket{\varphi_k} \ket{\alpha} = \sqrt{p_k}\ket{\varphi_{k+1}},
\end{align}
where the normalization constant $p_k$ is the probability of avoiding detection in the $k$-th step and 
\begin{align}
	W & = \bra{\alpha}P\ket{\alpha} e^{-i\delta\sigma_x}+\bra\alpha P^\perp\ket{\alpha}  e^{i\delta\sigma_x} \nonumber\\
	& = (\cos \delta) \bra{\alpha}(P + P^\perp) \ket{\alpha}  \ii 
		- i (\sin \delta) \bra{\alpha}(P - P^\perp) \ket{\alpha}  \sigma_x \nonumber\\
	& = (\cos \delta)\ii - i(\sin \delta)\langle A \rangle\sigma_x,
	\label{Wdef}
\end{align}
as $P+P^\perp = \ii$ and $\langle A \rangle = \bra{\alpha}A\ket{\alpha} = \bra{\alpha}P-P^\perp\ket{\alpha}$, recalling \ref{Adef}.
The matrix $W$ has eigenvalues $\lambda_{\mp} = \cos \delta\mp i \langle A \rangle \sin \delta$ and eigenstates $\ket{+},\ket{-}$. 

Because the above holds for any $k$, we have
\begin{align}
	W^N\ket{\varphi_0} = \left(
		\prod_{k=0}^{N-1}\sqrt{p_k}
		\right)\ket{\varphi_{N}}
	= \sqrt{p_{pass}}\ket{\varphi_{N}},
\label{eq:p_pass}
\end{align}
where $p_{pass}$ is the probability to pass all $N$ validation steps.
Let us look at what this becomes for large $N$, recalling $\delta = \frac{c}{N}$.
\begin{align}
	\lambda_{\mp}^N &= \left(\cos(\delta)\mp i \sin(\delta)\langle A \rangle\right)^N 
	= \left(e^{\mp i \delta \langle A \rangle} + O\left(\delta^{2}\right)\right)^N
	= \left(e^{\mp i \delta \langle A \rangle} \left(1 + O\left(\delta^{2}\right)\right)\right)^N \\
	& = e^{\mp i N \delta \langle A \rangle} \left(1 + N \times O\left(\delta^{2}\right) \right)  
	 = e^{\mp i c \langle A \rangle}  + O\left(N^{-1} \right).\label{eq:lambda_plusminus}
\end{align}
When we choose $N$ to be large (thus, small $\delta$), we get
\begin{align}
 	W^N &= e^{-i c\langle A \rangle\sigma_x}+O\left(\frac{1}{N}\right), 
\end{align}
meaning that we have approximately rotated the probe system by an amount proportional to $\langle A \rangle$.
Furthermore, 
\begin{align}
	\sqrt{p_{pass}} \ket{\varphi_{N}} & = e^{-ic \langle A\rangle\sigma_x}\ket{\varphi_0}+O\left(\frac{1}{N}\right)\ket{\tilde\varphi},
\end{align}
with $\ket{\tilde\varphi}$ some  normalized state.
By Eq.~\eqref{eq:p_pass} 
this means the probability that we failed some validation is small:
\begin{align}
	p_{pass}=1-O\left(\frac{1}{N}\right),
	\end{align}
and the final state can be also rewritten as 
\begin{align}
	\ket{\varphi_N} = e^{-i c \langle A \rangle\sigma_x}\ket{\varphi_0} + 
	O\left(\frac{1}{N}\right) \ket{\varphi'},
\end{align}
where $\ket{\varphi'}$ is some normalized state. 
We have demonstrated that we can apply the transformation $W^N$ with a low probability of failure
and high precision, i.e. $f=O(1/N)$ and $\eta=O(1/N)$. To prove Lemma \ref{lemma:protmeas} we set $c=\frac{\pi}{8}$. 

\subsection{A simple protective measurement: Identifying the four Wiesner money states}
\label{sec:simple}
Before describing the solution to the problem of protective tomography we give a simple application of the previous subsection for the case of the four Wiesner states.

To identify of the four states $\{\ket{0},\ket{1},\ket+,\ket-\}$ we can use a slightly modified version of the procedure described above. 
We choose $A=\sigma_x$, $c=\frac{\pi}{2}$, $\ket{\varphi_0}=\ket{0}$
and recall $\bra{0}\sigma_x\ket{0}=\bra{1}\sigma_x\ket{1}=0$ and 
$\bra{+}\sigma_x\ket{+}=-\bra{-}\sigma_x\ket{-}=1$,

If the money qubit $\ket{\alpha}$ was initially $\ket{+}$ or $\ket{-}$, the operation $W$ in Eq.~\eqref{Wdef} is exactly $e^{\mp i \delta \sigma_x}$. It is unitary, which means we never fail a verification.
The final probe state in these two cases will be $W^N \ket{0} = \mp i \ket{1}$.

On the other hand, if the money state is $\ket{0}$ or $\ket{1}$, 
then  $W$  is approximately the identity and  the probe will remain close to $\ket{0}$. 

Thus, when we measure the probe after $N$ rounds of interaction/verification in the computational basis, we find out whether the unknown state is an $x$-basis or $z$-basis state, allowing us to uniquely identify $\ket{\alpha}$ by measuring it in this basis.

\subsection{Proof of Lemma~\ref{le:sfp}: Single copy estimation of an expectation value.}
\label{sec:tomography}

The procedure for estimating the expectation value $\langle A \rangle$ is based on running the protocol in Sec.~\ref{sec:weak} for $m=336 \ln(2 \eta^{-1}) \nu^{-2}$ times, with $N=m/f$. We assume, w.l.o.g. that the precision parameter $\nu$ is smaller than some universal constant (if not, replace $\nu$ with that constant, and the result is improved).  We can already see using Lemma \ref{lemma:protmeas} that the total running time is $mN = O(\ln^{2}(\eta^{-1}) \nu^{-4} f^{-1})$, and that the overall failure probability is $O(\frac{m}{N})= O(f)$.  By applying the protective measurement procedure successfully $m$ times we obtain $m$ copies of the state 
\begin{equation} 
	\ket{\varphi_N} = \cos\left(\frac{\pi}{8}\langle A \rangle \right)\ket{0}
	- i\sin\left(\frac{\pi}{8}\langle A \rangle \right)\ket{1}
	+ O\left(\frac{1}{N}\right)\ket{\varphi'}, \label{phiN}
\end{equation} 
with some unknown normalized (error) state $\ket{\varphi'}$.
We can collect the desired statistics for estimating $\langle A \rangle$ by measuring each of   these $m$ copies in the $\sigma_y$ basis. Let  $\bar{p}_{y_+}$ be the  probability for getting the result $+1$ and   $p_{y_+}^{(m)}$ be the empirical frequency of a $+1$ result in these $m$ measurements.  

Using $\bra{y_+}=\frac{1}{\sqrt{2}}[\bra{0}-i\bra{1}]$, we get 

\begin{align}
	\bar{p}_{y_+} = \left|\braket{y_+}{\varphi_N}\right|^2
	& = \frac{1}{2}\left| \cos\left(\frac{\pi}{8}\langle A \rangle \right)
	   	- \sin\left(\frac{\pi}{8}\langle A \rangle\right) 
		+ O\left(\frac{1}{N}\right)              \right|^2 \\
	& =  \frac{1}{2} \left(1 - 2 \sin\left(\frac{\pi}{8}\langle A \rangle\right) 
		 \cos\left(\frac{\pi}{8}\langle A \rangle\right) \right)  + O\left(\frac{1}{N}\right)  \\
& = \frac{1}{2} \left( 1 - 
		\sin\left(\frac{\pi}{4}\langle A \rangle\right) 
		\right)  
	+O\left(\frac{1}{N}\right) \label{eq:p_y_plus}.
\end{align}

Next, we will show that 
we can estimate $\langle A \rangle$  by  $\frac{4}{\pi} \arcsin\left( 1 - 2p_{y+}^{(m)}\right)$,  with precision $\nu$ and confidence  $1-\eta$.

\begin{theorem}[Chernoff bound, see, e.g. {\cite[p. 66]{mitzenmacher2005probability}}] 
Let $X_1,\ldots,X_m$ be independent Bernoulli trials, $X=\frac{1}{m}\sum_{i=1}^m X_i$, $\mu=\mathbb{E}[X]$. For $0<\tilde\nu<1$, 
\[ \Pr(|X-\mu|\geq \tilde\nu \mu) \leq 2 \exp\left(-\frac{\tilde\nu^2 m \mu}{3}\right). \]
\end{theorem}
Let us choose $\tilde\nu=\frac{\nu}{4}$ and recall that
$m= \frac{336 \ln(2/\eta)}{\nu^2}$. Since  $|\langle A \rangle| \le 1$, the expectation value $\bar{p}_{y_+}$ is well bounded away from zero: $ \bar{p}_{y_+} \geq \frac{1}{2} - \frac{1}{\sqrt{8}}+O\left(\frac{1}{N}\right) \geq \frac{1}{7}$. Therefore,
\[ m= \frac{336 \ln(2/\eta)}{\nu^2} \geq \frac{3 \ln(2/\eta)}{\tilde\nu^2\bar p_{y+}}\]
By the Chernoff bound, the probability that $\left|p_{y_+}^{(m)}-\bar{p}_{y_+}\right| \geq \tilde \nu \bar{p}_{y_+} $ is at most 
\begin{align*}
2 \exp\left(-\frac{\tilde\nu^2 m \bar p_{y+}}{3}\right) \leq \eta.
\end{align*}

Our next goal is to show that we can estimate $\langle A \rangle$ easily and well using $p_{y_+}^{(m)}$.
We know that with confidence $1-\eta$, the value
$p^{(m)}_{y_+}$is within $\frac{\nu}{4} \bar{p}_{y_+} \leq \frac{\nu}{4}$ of $\bar{p}_{y_+}=\frac{1}{2} \left( 1 - 
		\sin\left(\frac{\pi}{4}\langle A \rangle\right) 
		\right)  
	+O\left(\frac{1}{N}\right)$  (see Eq.~\eqref{eq:p_y_plus}) .
	Moving the terms around, we get that
$\sin\left(\frac{\pi}{4}\langle A \rangle\right)$ is within 
$\frac{\nu}{2} + O\left(\frac{1}{N}\right)$ of $1-2p^{(m)}_{y_+}$, or that
\begin{align}
 \arcsin\left(1-2p^{(m)}_{y_+} - \frac{\nu}{2} - O\left(\frac{1}{N}\right)\right)
\leq \frac{\pi}{4}\langle A \rangle \leq \arcsin\left(1-2p^{(m)}_{y_+} + \frac{\nu}{2} + O\left(\frac{1}{N}\right)\right),
\end{align}
with confidence $1-\eta$. 

How far can $\frac{\pi}{4}\langle A \rangle$ be from $\arcsin\left(1-2p^{(m)}_{y_+}\right)$? We can use the Taylor series expansion $\arcsin(x + \delta) = \arcsin(x) + \frac{\delta}{\sqrt{1-x^2}} + O(\delta^2)$
for $x=1-2p_{y_+}^{(m)}$ and $\delta = \frac{\nu}{2} +O\left(\frac{1}{N}\right)$, assuming small $\nu$ and large $N$.
It now proves useful that we chose our procedure so that $\bar{p}_{y_+}$ in \eqref{eq:p_y_plus} is in the range $\frac{1}{2}-\frac{1}-{\sqrt{8}}-O(\frac{1}{N})\le\bar{p}_{y_+}\le \frac{1}{2}+\frac{1}{\sqrt{8}}+O(\frac{1}{N})$. 
Because of this, for reasonably small $\nu$ and big enough $N$, we can bound $|x|=\left|1-2p^{(m)}_{y_+}\right| \leq \frac{1}{\sqrt{2}} + O(\nu) + O\left(\frac{1}{N}\right) \leq \frac{3}{4}$, and $\frac{1}{\sqrt{1-x^2}} \leq \frac{4}{\sqrt{7}}$.
For small enough $\delta$, we can have the $O(\delta^2)$ term in the Taylor series smaller in magnitude than $\frac{|\delta|}{20}$.\footnote{\label{fn:universal_constant}This is guaranteed by the assumption we made in the beginning of the proof that $\nu$ is smaller than some universal constant.}
We can then conclude that with probability $1-\eta$, the value of $\frac{\pi}{4}\langle A \rangle$ is within $\left(\frac{4}{\sqrt{7}}+\frac{1}{20}\right)\delta \leq 0.781 \nu$ (see Footnote~\ref{fn:universal_constant})  of $\arcsin\left(1-2p_{y_+}^{(m)}\right)$, or alternatively,
\begin{equation}
\left| \langle A \rangle  - \frac{4}{\pi} \arcsin\left(1-2p_{y_+}^{(m)}\right) \right| \leq \nu.
\label{eq:approx_A}
\end{equation}

This concludes our proof of Lemma~\ref{le:sfp}. 
 
\subsection{Proof of Theorem \ref{theorem:prottomo}: tomography of a single qubit\label{sec:P5}}
 


Using Lemma~\ref{le:sfp}, let $\langle \tilde \sigma_j \rangle$ for $j\in \{x,y,z\}$ be the approximated expectation value  of each of the three Pauli operators  $\langle \sigma_j \rangle=\bra{\alpha}\sigma_j\ket{\alpha}$, with precision parameters $\tilde\nu=\epsilon/6,\ \tilde\eta=\eta/3, \tilde f=f$. The running time for getting all three values   is $3 \cdot O\left(\tilde f^{-1}\tilde\nu^{-4}\ln^2(\tilde\eta^{-1})  \right) = O\left(f^{-1}\epsilon^{-4}\ln^2(\eta^{-1})  \right)$ as required and the failure probability is $f\le 3\tilde f$ by the union bound.    
Conditioned that there were no failures, by using the union bound again, we get the required confidence value \begin{equation}
\Pr\left(\bigcap_{j \in \{x, y, z\}} |\langle \tilde \sigma_j \rangle - \langle \sigma_j \rangle |\leq \tilde\nu\right)\geq 1-3 \tilde\eta = 1-\eta.
\label{eq:nu_and_eta}
\end{equation}

To finish the proof, all we need is to show how, in the event there were no failures, we can construct  the state $\rho$ from the  three approximate expectation values $\langle \tilde \sigma_j \rangle$ such that $\rho$ is an approximation  of  $\ket{\alpha}\bra{\alpha}=\ii/2 + \sum_{j \in \{x,y,z\} }  \langle {\sigma}_j \rangle \sigma_j$ with fidelity at least $1-\epsilon$ and confidence at least $1-\eta$. 
Let $\tilde\rho=\ii/2 + \sum_{j \in \{x,y,z\} }  \langle \tilde{\sigma}_j \rangle \sigma_j$. This Hermitian trace one operator is  not necessarily positive semidefinite. We therefore choose $\rho$  to be  the closest state to $\tilde\rho$ , that is  $\rho=\argmin_{\tau} D(\tilde\rho,\tau)$, where $\tau$ runs over all single qubit mixed states and $D(.,.)$ is the trace distance $D(\alpha,\beta)=\frac{1}{2} ||\alpha - \beta||_{tr}$ and $||A||_{tr} = \tr(\sqrt{A A^\dagger})$). 
Using the triangle inequality and the definition of $\rho$, we obtain
\begin{equation}
D(\rho,\ket{\alpha}\bra{\alpha})\le D(\rho, \tilde\rho) + D(\tilde\rho,\ket{\alpha}\bra{\alpha})\le 2 D(\tilde\rho,\ket{\alpha}\bra{\alpha}).
\label{eq:D_rho_alpha}
\end{equation}
 Then the fidelity of the final state satisfies that with probability at least $1-\eta$,
\begin{align}
F(\rho,\ketbra{\alpha}) &\geq 1-D(\rho,\ketbra{\alpha}) \label{eq:D_rho_i_alph_i}\\
&\geq 1-2 D(\tilde \rho,\ketbra{\alpha})\\
&\geq 1-2\sum_{j\in \{x,y,z\} }\frac{1}{2} \norm{
			\left(\langle \sigma_j \rangle  - \langle \tilde \sigma_j\rangle \right) \sigma_j 
			}_{tr}  \\
&= 1-2 \sum_{j\in \{x,y,z\} } \left|\langle \sigma_j \rangle  - \langle \tilde \sigma_j \rangle\right| 
\geq 1 -6 \tilde\nu  = 1- \epsilon,
\end{align}
where in the first step, we used the fact that if one of the states $\alpha$ or $\beta$ is pure, then $F(\alpha,\beta) \geq 1- D(\alpha,\beta)$
; in the second step we used Eq.~\eqref{eq:D_rho_alpha}; in the third step  we used the triangle inequality for the trace norm; we used Eq.~\eqref{eq:nu_and_eta} in the last inequality; and the definition of $\tilde\nu=\epsilon/6$ in the last step. The properties of the trace distance which we used are shown, for example, in the  textbook~\cite{nielsen2010quantum}.

This completes the proof of Theorem~\ref{theorem:prottomo}.

\subsection{Proof of Corollary \ref{cor:protective_tomography_n_qubits}: Protective tomography of $n$ qubits.\label{sec:PTn}}
We use Theorem~\ref{theorem:prottomo} to apply tomography to each of the $n$ qubits, with parameters $\tilde\epsilon=\epsilon/n$, $\tilde \eta = \eta / n$, $\tilde f = f/n$. By the union bound, the failure probability is at most $n \tilde f =f$,  and the error probability is at most $n \tilde \epsilon = \epsilon$, as required. Let $\rho = \rho_1 \otimes \ldots \otimes \rho_n$, where $\rho_i$ is the $\tilde \epsilon$ approximation of $\ket{\alpha_i}$, as provided by the theorem.  

\begin{equation*}
F(\rho,\ketbra{\alpha}) = \prod_{i=1}^n F(\rho_i,\ketbra{\alpha_i}) \geq  (1-\tilde \epsilon)^n \geq  1-n\tilde \epsilon=1-\epsilon
\end{equation*}
 where the first step follows from $F(\alpha \tensor \beta, \gamma \tensor \delta)=F(\alpha,\gamma)F(\beta,\delta)$.
 
 By applying this procedure directly, we get a total running time of $n \cdot O\left(n^5 f^{-1}\epsilon^{-4}\ln^2(n \eta^{-1})  \right)$. But we can save a factor of $n$ to get a total running time of $O\left(n^5 f^{-1}\epsilon^{-4}\ln^2(n \eta^{-1})  \right)$, as required, by running the procedures in parallel, using the same idea that was explained in the previous section, see the analysis preceding Eq.~\eqref{eq:verifications_in_parallel}. This completes the proof of Corollay~\ref{cor:protective_tomography_n_qubits}.
\section{Discussion: Comparing the two attacks}
\label{sec:discussion}

We presented two different attacks for counterfeiting quantum money in the strict testing regime. Both attacks are based on known methods to exploit the quantum Zeno effect in order to learn a quantum state without disturbing it. 
Despite their similarities, the methods are conceptually different and are not always interchangeable. In Appendix~\ref{sec:general}, we discuss a generalized version of the BT attack and show that it fails in the most general scenario. There, we have to choose the PM attack.

One may ask why (or whether) the PM attack  is less useful than BT in other scenarios. 
The first possible advantage of the BT attack is that it identifies a state from a given list of states, instead of producing only an estimate of the state. 
However, a simple modification of the PM protocol can also identify the state, in a similar fashion as the BT attack (just like the PM attack in Section~\ref{sec:simple}). Another advantage might be in terms of resources.  
Although our analysis in both cases is not necessarily optimal, in all the cases we checked the PM attack (or simple adaptations of it) does not use more queries than the BT attack. To conclude, our analysis suggests that the PM attack has both qualitative and quantitative advantages compared to the BT attack.  

One way to gain efficiency may be to use phase estimation on the transformation for $W$ in Eq. \eqref{Wdef}, which is approximately a rotation. We could improve the efficiency even further by using multiple probes per qubit of the money state. 
The two attacks also lead to different behaviors of the probe qubit: in the BT attack, the probe qubit moves only if the money state lies in a very narrow window of angles around the reflected state (see Fig.~\ref{fig:recurrence_relation_plot}); in the PM attack, the probe qubit rotates by an angle proportional to the expectation value that we want to measure.

In both attacks, we have two systems, the probe and the money. The fundamental idea is to cause a minimal kick to the money state in each iteration. Let us  use $Q_i$ to describe the completely positive trace preserving channel acting on the money state in the $i$th iteration and $V_\$$ to indicate the bank's verification step  that follows $Q_i$. Ideally, we want $Q_i(\ket\$\bra\$)\approx\ket\$\bra\$ $ for all $i$. This could happen for one of two reasons, either $\ket\$\bra\$$ is a fixed state for the channels $Q_i$, or $Q_i\approx \ii$ (this is what we call a weak channel). In general, the channel is not fixed, since it depends on the state of the pointer. For both methods, the initial input state is unknown, and therefore the channel has to be initially weak. Does it remain this way? 

In the case of the PM attack, the interaction term is weak, so the corresponding channel is also always weak. Moreover, the attack is set up so that the channel is invariant in time to a good approximation. The BT attack is more interesting. Initially, the channel is weak, because the pointer is very close to the $\ket{0}$ state. For most input states, the pointer stays very close to its initial position (see detailed analysis in Appendix~\ref{sec:general}). However, for a small family of states around one of the fixed points, the pointer starts to move towards the state $\ket{1}$ and the channel becomes stronger. This strange behavior of the channel that goes from weak to strong as a result of different input states has not been previously noticed, as far as we know. We believe that further insight into these types of measurements can lead to interesting quantum information protocols beyond those mentioned here.

\section{Acknowledgments}
We would like to thank Scott Aaronson, Dorit Aharonov, David Gosset, Guy Kindler, Carl Miller,  Lev Vaidman and Stephen Wiesner for valuable discussions, and the Simons institute Quantum Hamiltonian Complexity program, during which a substantial part of this work was done.
DN has received funding from the People Programme (Marie Curie Actions) EU's 7th Framework Programme under REA grant agreement No. 609427, and this research has been further co-funded by the Slovak Academy of Sciences.
He also thanks the Slovak Research and Development Agency grant APVV-0808-12 QIMABOS. 
OS also thanks the ARO Grant W922NF-09-1-0440 and NSF Grant CCF-0905626.
DU  was supported by  the Estonian ICT  program 2011-2015
(3.2.1201.13-0022), the  European Union through the  European Regional
Development Fund through the  sub-measure ``Supporting the development
of R\&D of info and communication technology'', by the European Social
Fund's Doctoral  Studies and  Internationalisation Programme  DoRa, and by
the Estonian Centre of Excellence  in Computer Science, EXCS.  AB is  supported by NSERC, Industry Canada and CIFAR.
\appendix
\section{The basic bomb-tester and semi-faulty bombs}
\label{sec:general}
Let us now discuss the BT attack of Section~\ref{sec:attack}, and demonstrate that it works efficiently and safely for a given list of hidden states, but stops working when the secret states are completely unknown, or chosen too close to each other. 

\subsection{Using a longer list of secret states is still unsecure against the bomb-testing adaptive attack}
\label{testLIST}

Counterfeiting quantum banknotes with arbitrary secret single-qubit states becomes
more complicated.
Let us continue what we did before and try to identify an unknown qubit state $\ket{\alpha}$, if we have access to a strict validation procedure (see Figure~\ref{fig:bomb}).
Our approach is to try to find a unitary $R$ which acts as $R\ket{\alpha} = \ket{\alpha}$. 

Let us assume that the unknown state comes from a list of possible states $\ket{\alpha} \in \mathcal S$. 
We can just pick one of the states $\ket{\beta}$ from this list, and use our previous attack
with the controlled reflection $R = 2\ket{\beta}\bra{\beta}-\ii$ instead of the controlled $X$ (CNOT) in Figure~\ref{fig:attack}.

First, if the unknown state $\ket{\alpha}$ is exactly our tested $\ket{\beta}$, we have the ``dud case'', which we identify without fail by measuring $\ket{1}$ on the first register (as $R\ket{\alpha} = \ket{\alpha}$). 

Second, if $\ket{\alpha}\neq \ket{\beta}$, the operation $R$ does ``something'' to $\ket{\alpha}$. We can choose the phases of the vectors involved to give us a real angle $0\leq \theta \leq \frac{\pi}{2}$ with $\cos\theta=\braket{\alpha}{\beta}$ and
\begin{align}
	R \ket{\alpha} = \cos(2\theta) \ket{\alpha} + \sin(2\theta) \ket{\alpha^{\perp}}.
\end{align}
Let us now look at our testing procedure in Figure~\ref{fig:attack}.
If we haven't been caught after some number $k$ of rounds, 
the control register contains a state parametrized by an 
angle $\varphi_k$, i.e. $\ket{\varphi_k} \propto (\cos \varphi_k) \ket{0}+ (\sin \varphi_k) \ket{1}$.
We have the valid quantum money state in the second register. After rotating the first register by $\delta$, the state of the system becomes
\begin{align}
	\left( \cos (\varphi_k+\delta) \ket{0} + \sin (\varphi_k+\delta) \ket{1} \right) 
	\,\ket{\alpha}.
\end{align}
We apply the controlled probe $C_R$ (instead of CNOT) to both registers, and obtain
\begin{align}
	\cos (\varphi_k+\delta) \ket{0}\ket{\alpha} + \cos(2\theta) \sin (\varphi_k+\delta) \ket{1}\ket{\alpha}
	+ \sin(2\theta) \sin (\varphi_k+\delta)  \ket{1}\ket{\alpha^\perp}.
\end{align}
We now measure the second register. The probability of being caught in this round is
$\sin^2(2\theta) \sin^2 (\varphi_k+\delta)$. 
More importantly, after a successful test, we are left with the unnormalized state
\begin{align}
	\left( \cos (\varphi_k+\delta) \ket{0} + \cos(2\theta) \sin (\varphi_k+\delta) \ket{1}
	\right) \, \ket{\alpha}.
\end{align}
It means that the state of the first register as a two component (unnormalized) vector is transformed as
\begin{align}
	\ket{\varphi_{k+1}}  
	= \mmm{1}{0}{0}{\cos(2\theta)} R_{\delta} \ket{\varphi_k}
	= \underbrace{\mmm{\cos \delta}{-\sin \delta}{q\sin \delta}{ q\cos \delta}}_{T} \ket{\varphi_k}
	=T \ket{\varphi_k}, \label{Tdef}
\end{align}
where we labeled 
\begin{align}
	q = \cos (2\theta). \label{Qdef}
\end{align}
When we pass all the tests, the unnormalized state of the first register at the end of the protocol is $\ket{\varphi_N} = T^N \ket{0}$. 
We can get the probability of passing all the tests by taking the norm squared of this vector.

There are two simple special cases.
First, when we guess correctly, we have $\theta=0$ ($q=1$), which gives us $T=R_{\delta}$ and $\varphi_{k+1}=\varphi_k+\delta$ (the ``dud'' case). Then $\bra{1}T^N\ket{0}=1$, and in $N$ steps the first qubit is rotated to $\ket{1}$, plus we cannot get  caught as since we are  doing nothing to the banknote.
Second, for $\theta = \frac{\pi}{2}$ ($q=-1$), we have $T^2 = \ii$, and the system behaves like the $\ket{-}$ case in Section~\ref{sec:attack}, with the first register remaining in the state $\ket{0}$ after an even number of rounds, i.e. $\bra{0}T^N\ket{0}=1$. We are also never caught doing anything illegal.

Third, we have a much more interesting general case $\theta_{min} \le \theta < \frac{\pi}{2}$, 
meaning $|q|<1$. 
We claim that we will pass all the tests successfully, and measure $\ket{0}$ in the first register at the end, with probability close to 1.
We are choosing a large $N$, which means small $\delta$, for which we can write
\begin{align}
		T = \mmm{1}{-\delta}{q\delta}{q} + \Delta T,
		\label{T2}
\end{align}
with a small error term $\norm{\Delta T}=O(\delta^2)$. 
Let us now calculate the state of the first register after $N$ rounds using \eqref{T2}, keeping track of the errors we accumulate:
\begin{align}
		T^N \ket{0} &= T^N \vvv{1}{0}
		= T^{N-1} \vvv{1}{\delta q} 
			+ \underbrace{T^{N-1}\Delta T \vvv{1}{0}}_{\ket{v_1}} \nonumber\\
		&= T^{N-2} \vvv{1}{\delta \left(q+q^2\right)} 
				+ \underbrace{T^{N-2}\left(\vvv{- \delta^2 q}{0} + \Delta T \vvv{1}{\delta q}\right)}_{\ket{v_2}} + \ket{v_1} \nonumber\\
		&= T^{N-3} \vvv{1}{\delta \left(q+q^2+q^3\right)} 
				+ \underbrace{T^{N-3}\left(\vvv{- \delta^2 \left(q+q^2\right)}{0} + \Delta T \vvv{1}{\delta \left(q+q^2\right)}\right)}_{\ket{v_3}} + \ket{v_2} + \ket{v_1} \nonumber\\
		&\,\,\,\,\vdots \nonumber\\
		&= \vvv{1}{\delta \left(q+q^2+\dots + q^N\right)} 
				+	\ket{v_N} + \dots + \ket{v_1}, 
		\label{finalstate2}
\end{align}
where the error vectors and their norms are
\begin{align}
		\ket{v_k} &= T^{N-k} \left(\vvv{-\delta^2\left(q+q^2+\dots+q^{k-1}\right)}{0}+ \Delta T \vvv{1}{\delta\left(q+q^2+\dots+q^{k-1}\right)}\right),\\
		\norm{\ket{v_k}} &\leq O\left(\delta^2\left(1+q+q^2+\dots+q^{k-1}\right)\right) \leq O\left(\frac{\delta^2}{1-q}\right).
\end{align}
because
we have $\norm{T}\leq 1$ in \eqref{Tdef} implying $\norm{T^k \ket{w}} \leq \norm{\ket{w}}$ for any $k\geq 1$, as well as $\norm{\Delta T}=O\left(\delta^2\right)$.
We can now look at \eqref{finalstate2} and lower bound the amplitude 
\begin{align}\label{eq:0TN0}
		\bra{0} T^N \ket{0} \geq  1 - N \norm{\ket{v_N}} 
		\geq1- O\left(  \frac{N\delta^2}{1-q} \right) 
		\geq 
		1-O\left( N^{-1}\theta_{min}^{-2} \right),
\end{align}
using $1-q = 1-\cos(2\theta) = \Omega\left(\theta^2\right)$ 
for small $\theta$.
Therefore, there exists a constant $c$ such that when we choose 
\begin{align}	\label{eq:BT_O}
		N=c f^{-1}  \theta_{min}^{-2} = O \left( f^{-1}  \theta_{min}^{-2}\right),
\end{align}
the probability of passing all $N$ tests while measuring $\ket{0}$ at the end will be
\begin{align}
	\left|\bra{0} T^N \ket{0}\right|^2 \geq  1 - f.
\end{align}

This procedure rules out (or identifies) $\ket{\beta}$, one of the possible states from $\mathcal{S}$. We can now proceed to eliminate or identify further states, also getting rid of the pesky smallest possible angles $\theta_{min}$ which make us use large $N$, so that the procedure gets more efficient later on. 

However, in general, in the worst case, in order to identify a single qubit, we need to repeat this procedure $r:=\lvert\mathcal S\rvert$ times. 
Moreover, we need to identify all $n$ qubits. We can do it safely by choosing a much smaller $f' = f / (nr)$,
resulting in $N'= O\left( n r \theta_{min}^{-2} f^{-1} \right)$, which makes the entire attack succeed with probability $1-O\left(f\right)$. The total number of queries in the attack is at most $n \times r \times N'=O\left(n^2 r^2 \theta_{min}^{-2} f^{-1}\right)$.
However, just as in Section~\ref{sec:attack}, 
we can do the attack in parallel, modifying all $n$ banknote qubits slightly (using $n$ individual probe qubits)
and submitting them for verification. In this case, we save a factor of $n$, and will require only
$O\left(n r^2 \theta_{min}^{-2} f^{-1}\right)$ rounds of verification.

\subsection{The bomb testing attacks does not work for general unknown states}
\label{bomb_testing_fails}

Let us relax our assumptions and look at what happens if $\theta_{min}$ is not bounded from below, i.e. if the states that we are receiving are completely unknown, or perhaps from a continuous range of states. We run into a new problem, because what we do can no longer always be interpreted as a weak measurement. Let us illustrate this point.

When the set of possible money states is dense, there is no minimal angle between two possible states. Similarly, when we do not know what the state could be, we can only guess, and attempt to do a controlled reflection about a random axis. 
Thus, the counterfeiter has to choose some $\theta_{min}$ since the other parameters $N,\delta$ explicitly depend on $\theta_{min}$.  This introduces a fourth option to the list above: $0<\theta<\theta_{min}$. 

\begin{figure}
\begin{center}
\includegraphics[width=11cm]{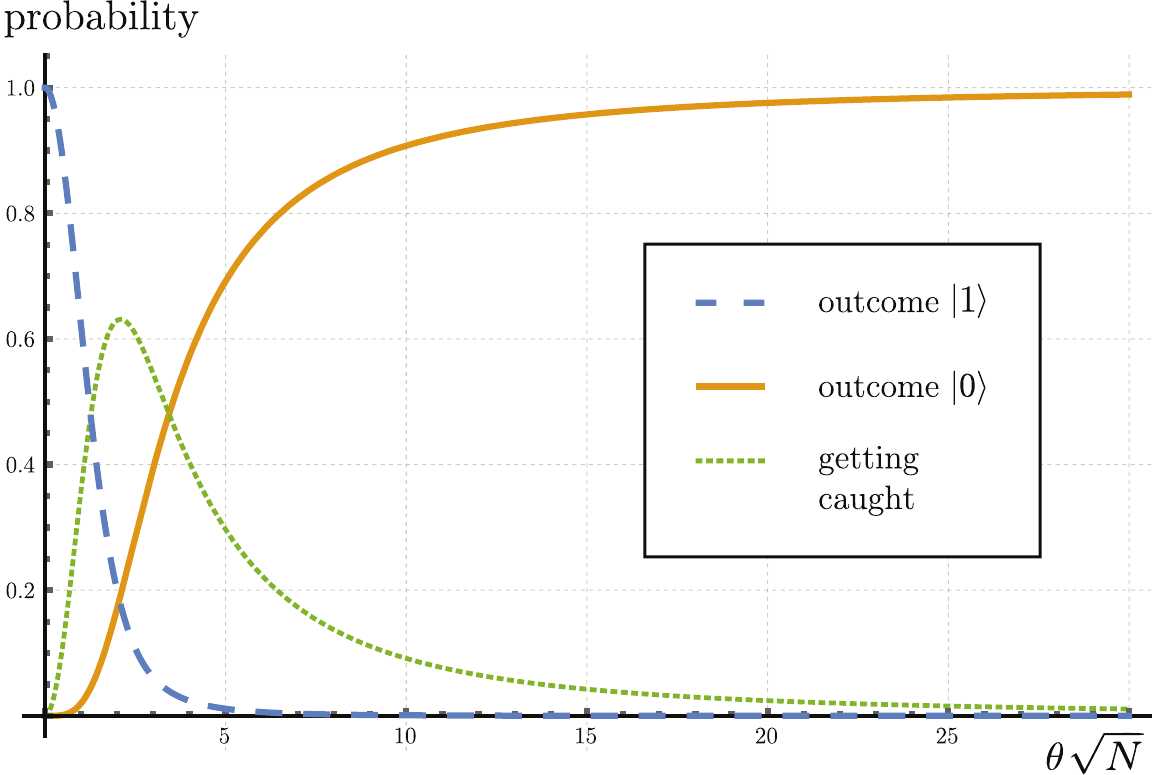}%
\caption{The probability of each of the outcomes of the protocol from Section~\ref{testLIST} as a function of $\theta$. Note that the x-axis is $\theta \sqrt N$.  
The probability of an getting caught by the bank, shown in the dotted green line, goes to 0 as $\theta \sqrt N \rightarrow \infty$, as well as with $\theta\sqrt{N}\rightarrow 0$. 
}%
\label{fig:recurrence_relation_plot}%
\end{center}
\end{figure}

Before explaining \emph{why}, we first explain \emph{what} happens.
Using Mathematica, we analytically computed the outcome of the protocol from Section~\ref{testLIST}, as a function of $\theta \sqrt{N}$, in the limit $N \rightarrow \infty$, which is depicted in Fig.~\ref{fig:recurrence_relation_plot}. 
 It can be seen that we do not get caught with high probability only if either $\theta \ll \frac{1}{\sqrt N}$ (in which case, the outcome of the measurement of the probe qubit is $\ket{1}$) or if $\theta \gg \frac{1}{\sqrt N}$ (in which case, the outcome of the measurement of the probe qubit is $\ket{0}$). If $\theta$ is in the order of  $\frac{1}{\sqrt N  }$ there is a constant probability of getting caught by the bank. If we choose a random state $\ket{\beta}$ that we want to identify, the probability to land in the safe region $\theta \ll \frac{1}{\sqrt N}$ (which allows us to identify the money state as close to $\ket{\beta}$), is much smaller than the probability of landing in the dangerous zone $\theta = \Theta(\frac{1}{\sqrt N})$. We need many calls to the protocol, therefore, we cannot tolerate a constant probability of being caught. To conclude, independent of how $N$ is chosen, we will most likely get caught by the bank before getting an approximation to a single qubit of the bank note (and we need to succeed in that $n$ times). Therefore, the BT attack fails when the set of possible states is infinite or unknown.
 
The basic building block of the protocol we analyzed in Section~\ref{testLIST} is a rotation followed by an interaction, followed by validation. We want the validation to succeed with high probability. Prior to the controlled rotation (the interaction stage) the money state is the original $\ket{\alpha}$. What happens after the interaction? There are two cases: 
\begin{enumerate}
	\item The money state is an eigenstate of $R$ (i.e. $\theta\in\{0,\pi/2\}$), so the unknown system stays factorized and validation always succeeds. 
	\item $\theta\in (0,\pi/2)$. We can be assured of success (not being detected) when the interaction stage  is ``weak enough''. 
\end{enumerate}

The interaction is ``weak enough'' for our purposes whenever we do not disturb the system much, so that we remain undetected even as we test it many times. 
We want to safely run $O(N)$ rounds,
so it is enough if the probability 
to be caught in any particular round is $o\left(N^{-1}\right)$; clearly $O\left(\delta^2\right) = O\left(N^{-2}\right)$ is enough.

The critical case appears sneakily, when the state $\ket{\alpha}$ is almost, but not quite a $+1$ eigenstate of $R$, which means $q\lesssim 1$ in \eqref{Qdef}. In this case, $R$ does not disturb the unknown system much, and the transformation $T$ \eqref{T2} is approximately
\begin{align}
	T_{q\lesssim 1} \approx \mmm{\cos \delta}{-\sin \delta}{\sin \delta}{ \cos \delta},
\end{align}
resulting in a rotation of the state $\ket{\varphi_k}$ to $\ket{\varphi_k+\delta}$.
In this way, in a constant fraction of $N$ of rounds, our probe rotates to a state where its overlap with $\ket{1}$ becomes a constant, so we actually start to significantly apply $R_{q\lesssim 1}$ to the system. Although $R_{q\lesssim 1}$ does not disturb the system much, if $q = 1 - O\left(\sqrt{\delta}\right)$, the {\em cumulative} probability of detection can grow as we repeat many rounds; this is our ultimate doom. 
The the probe/system  would become
$\cos \varphi_k \ket{0} \ket{\psi} + \cos \varphi_k \cos \theta \ket{1} \ket{\psi}
+ \cos \varphi_k \sin \theta \ket{\psi^\perp}$,
so the probability of failure in this round is $\cos^2 \varphi_k \sin^2 \theta = O\left( \delta\right)$
for $\Theta\left(\varphi_k\right) = 1$.
Note that we can nicely avoid this problem when using the protective measurement method from Section~\ref{sec:weak}, ensuring a sufficient upper bound on the disturbance of the unknown system in any round.

Let us finish by saying why this is not a problem for $q$ well bounded away from $1$, in which case our interaction with the system can be always ``weak enough''. At the start of a particular round, the probe qubit is in the state $\ket{\varphi_k} = \cos \varphi_k \ket{0}+\sin \varphi_k\ket{1}$.
We then rotate it by $\delta$ to $\ket{\varphi_k+\delta}$.
Next, we perform the controlled reflection, run through a validation step and when we succeed, we end up in the unnormalized state \eqref{Tdef}.
Normalizing it, we find the angle $\varphi_k$ from the beginning of the round changed to $\varphi_{k+1}$, which for positive\footnote{For $q\leq 0$, we can bound $\left| \varphi_k \right| \leq \delta$, so the measurement is always weak.} $q$ is less than $\varphi_k+\delta$. It turns out that this iterative process has an upper bound. Moreover, for reasonable $q$ this upper bound is on the order of $\delta$, as can be learned from \eqref{finalstate2}. 
Thus, the probability of failure in any round is upper bounded by $O\left(\delta^2\right)$, and the measurement we do is weak enough for our purposes. 
We will not be detected in $N$ rounds with probability more than on the order of $\delta = O\left(\frac{1}{N}\right)$.

~\\
\bibliography{TomographyOnWiesnersMoneyQICresubmit}
\bibliographystyle{alphaabbr}
\end{document}